\documentclass[aps,twocolumn,superscriptaddress,showpacs,preprintnumbers,amsmath,amssymb,floatfix]{revtex4-1}

\usepackage{graphicx,tabularx}
\usepackage{dcolumn}
\usepackage{bm}
\usepackage{color}
\usepackage{epstopdf}
\usepackage{upgreek}
\usepackage{color}
\usepackage{array}
\usepackage{multirow}

\begin{document}
 

\title{Weak electron irradiation suppresses the anomalous  magnetization of N-doped diamond crystals }


\author{Annette Setzer}
\affiliation{Division of Superconductivity and Magnetism, Felix-Bloch Institute for Solid State
Physics, Universit\"at Leipzig, 04103 Leipzig, Germany}

\author{Pablo  D. Esquinazi}
\email{esquin@physik.uni-leipzig.de}
\affiliation{Division of Superconductivity and Magnetism, Felix-Bloch Institute for Solid State
Physics, Universit\"at Leipzig, 04103 Leipzig, Germany}

\author{Olesya Daikos}  
\affiliation{Leibniz Institute of Surface Engineering (IOM), 04318 Leipzig, Germany}
\author{Tom Scherzer}
\affiliation{Leibniz Institute of Surface Engineering (IOM), 04318 Leipzig, Germany}

\author{Andreas P\"oppl}
\affiliation{Division of NMR, Felix-Bloch Institute for Solid-state Physics, Universit\"at Leipzig, 04103 Leipzig, Germany}


\author{Robert Staacke }
\affiliation{Division of Applied Quantum Physics, Felix-Bloch Institute for Solid-state Physics, Universit\"at Leipzig, 04103 Leipzig, Germany}

\author{Tobias L\"uhmann }
\affiliation{Division of Applied Quantum Physics, Felix-Bloch Institute for Solid-state Physics, Universit\"at Leipzig, 04103 Leipzig, Germany}

\author{Sebastien Pezzagna}
\affiliation{Division of Applied Quantum Physics, Felix-Bloch Institute for Solid-state Physics, Universit\"at Leipzig, 04103 Leipzig, Germany}


\author{Wolfgang Knolle}
\affiliation{Leibniz Institute of Surface Engineering (IOM), 04318 Leipzig, Germany}

\author{Sergei Buga}
\affiliation{Technological Institute for Superhard and Novel Carbon Materials,
 Troitsk, Moscow, 108840 Russia}
 \affiliation{Moscow Institute of Physics and Technology,  Dolgoprudny, Moscow Region, 141701 Russia}

\author{Bernd Abel}
\affiliation{Leibniz Institute of Surface Engineering (IOM), 04318 Leipzig, Germany}

\author{Jan Meijer}
\affiliation{Division of Applied Quantum  Physics, Felix-Bloch Institute for Solid-state Physics, Universit\"at Leipzig, 04103 Leipzig, Germany}



\date{\today}
\begin{abstract}
  {Several
  diamond bulk crystals with  a    concentration of electrically neutral single substitutional nitrogen atoms of $\lesssim 80~$ppm,
   the so-called  C- or P1-centres,  were irradiated with electrons at 10~MeV energy and low fluence. The results show a 
   complete suppression of the irreversible behavior in field and temperature  of the magnetization below  $\sim 30$~K, after
  a decrease of  $\lesssim 40~$ppm in the concentration of  C-centres produced by the electron irradiation.  
  This result indicates that  magnetic C-centres  are at 
  the origin of the large  hysteretic behavior found recently in  nitrogen-doped diamond crystals. 
  This is remarkable because of the relatively low density of  C-centres, stressing the
extraordinary role of the C-centres in triggering those phenomena in diamond at relatively high temperatures. 
  After annealing the samples at high temperatures  in vacuum, the hysteretic behavior is partially recovered. }
\end{abstract}
\maketitle

\section{Introduction}
\label{introduction}
\subsection{Superconductivity in diamond: a still incomplete puzzle}\label{intro}
Superconductivity (SC) was recognized in doped diamond through  a relatively broad
transition in the electrical resistance at 2.5~K$~\ldots$ 2.8~K    
after doping it with $\sim 3\%$ boron \cite{EKI04,tak04}.  A critical temperature onset of 8.7~K was reported in
polycrystalline diamond thin films one year later \cite{tak05}. 
Angle-resolved photoemission spectroscopy (ARPES) studies revealed
that holes in the diamond bands determine the metallic character of the heavily
 boron-doped superconducting diamond \cite{yok05}.
Meanwhile, there are indications of  SC  below 12~K in
B-doped carbon nanotubes  \cite{mur08},  at 
$\sim 25~$K in heavily B-doped diamond
samples   \cite{oka15} or even at $\sim 55$~K in 27\% B-doped Q-carbon, an amorphous form  of carbon \cite{bha17}.
In spite of several experimental and theoretical studies  on the influence of boron as a  trigger of SC in diamond and in some 
carbon-based compounds, the real role of boron and the origin of SC in diamond  is not so clear as it may appear. For example, 
 scanning tunneling microscopy studies  of superconducting B-doped polycrystalline diamond 
samples revealed
granular SC with an 
 unclear  boron concentration within the superconducting regions \cite{zha14}. In spite of an apparently homogeneous
 boron concentration,  strong modulation of the order parameter were reported in \cite{zha14}, putting into question the
real role of boron in the SC of diamond. High resolution structural studies accompanied by electron energy loss spectroscopy  (EELS) 
on B-doped diamond single crystals revealed the presence of boron in distorted regions of the diamond lattice inducing
 the authors to argue that 
SC may appear in the disordered  structure and not in the defect-free B-doped lattice of diamond \cite{bla14}. 
This peculiarity may explain the high sensitivity of the superconducting
critical current between hydrogen- and oxygen-terminated boron-doped diamond \cite{yam17}.

Furthermore, we may ask whether the SC phase in B-doped diamond is homogeneously distributed in 
the reported samples. This does not appear to be in general the case. For example, the SC phase 
was observed only within a  near surface region
of less than $\sim 1~\mu$m in heavily B-doped diamond single crystals, showing clear polycrystallinity and granular properties \cite{blaepl}.
Some studies associated the Mott transition observed as a function of the B-concentration 
with metallic B-C bilayers; the measured SC was
detected at the surface of the doped crystals having a short modulation period between the B-C bilayers \cite{pol16}. 
Apparently, differences in the growth processes 
of the diamond samples cause  differences in the morphology  of the SC phase, an issue that still  needs more studies.

Several recent studies indicate a granular nature of the samples' structure and of the  SC phase \cite{pol16,zha11,zha16,zha19}. 
The granular nature of the SC phase and localized disorder 
may play an important role as recently reported experimental facts indicate, namely: (a) Not always there is  a clear 
correlation between
the B-concentration threshold for the metal-insulator transition 
and the one for SC \cite{bou17}, and (b) there is no simple dependence between
the free-carrier concentration and the critical temperature characterized by transport measurements \cite{cre18}. 

Another open issue in the SC puzzle of doped diamond is the fact that implanting boron into diamond
via irradiation does not trigger SC at least above 2~K \cite{hee08}. It has been argued that this irradiation
process does not trigger SC because of the produced defects that remained in the diamond lattice after
ion irradiation. However, this is not at all clear  because  the sample was heated to $900^\circ$C
during irradiation and afterwards annealed at $1700^\circ$C in vacuum \cite{hee08}, see also similar results in 
\cite{bev16,tsu12,tsu06}. It might be that the absence of SC in the 
irradiated samples after high temperature annealing is related to  the absence of certain defects and not other way around \cite{bas08}.
That  certain lattice defects are of importance for SC can be seen from 
the reentrance of SC observed in ion-irradiated B-doped diamond  after  high-temperature 
annealing \cite{cre18}. 
This reentrance of SC has been attributed to the partial removal of vacancies
previously produced by
light ion irradiation. It has been argued that vacancies  change the effective carrier density compensating the
boron acceptors and therefore suppressing SC  \cite{cre18}. 

\subsection{Defect-induced phenomena and the unconventional magnetization observed in nitrogen-doped diamond crystals}
According to recent studies, the reported SC  in B-doped diamond appears to be 
 more complicated because not only the boron concentration matters but also some kind of disorder or even  magnetic
order may play a role. 
 Coexistence of SC and ferromagnetism (FM) was recently reported in hydrogenated 
 B-doped nanodiamond films at temperatures below 3~K \cite{zha17,zha20}.  Earlier studies  
 \cite{tala05} showed the existence of ferromagnetic hysteresis 
 at room temperature in the magnetization  of  nanodiamonds after nitrogen or carbon irradiation.
Recently published studies revealed the existence of large hysteresis in field and temperature 
in the magnetization below an apparent critical temperature 
$T_c \simeq 30~$K  in B-free bulk  diamond single crystals. Those crystals were produced under high-temperature and high-pressure 
(HTHP) conditions with an average N-concentration $\lesssim 100$~ppm \cite{bardia}.
In this last work, a correlation was found between
the strength of the hysteretic behavior in the magnetization  with the N-content, in particular with the concentration of C-centres. 

We would like to stress that the possible origins for an irreversible behavior in field and/or temperature in the magnetization of a
material can be: (a) Intrinsic magnetic anisotropy, (b) the existence of magnetic domains and the pinning of their walls, or (c) the
  pinning of vortices or fluxons in a superconducting matrix. The origin (a) can be ruled out because the measured behavior
does not depend on the direction of the applied field with respect to the main axes of the crystals nor on their shape \cite{bardia}. This
independence has been confirmed once more in the diamond crystals studied in this work.  

The relatively 
low average N-content and further  details of the 
measured diamond samples  
suggest the existence of  a granular structure in the concentration of nitrogen \cite{bardia}. 
  
\begin{figure}
\includegraphics[width=1\columnwidth]{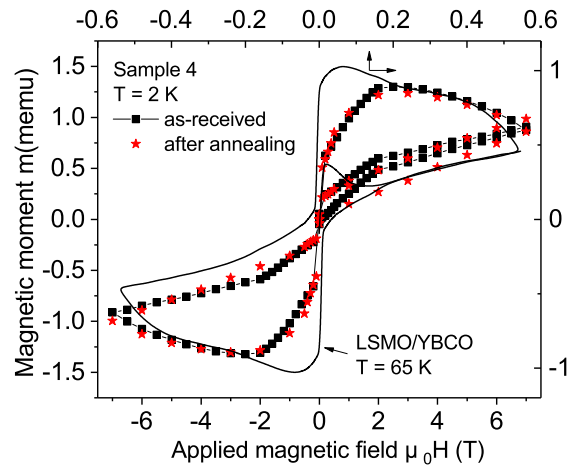}
\caption{\label{ht} Left-bottom axes: Magnetic field hysteresis  loop at $T = 2~$K of the magnetic moment of 
sample 4 studied in this work, before and after annealing four hours at 1200$^\circ$C in vacuum. 
Right-upper axes: ferromagnetic/superconducting LSMO/YBCO bilayer 
at 65~K taken from \cite{bardia}. 
A small diamagnetic linear field background was subtracted from the data points.}
\end{figure}

To facilitate  the comparison of the present results  with those of Ref.~\cite{bardia}, 
we give a summary below of the main results obtained  from the measurements of 
eight different nitrogen doped crystals:\\ 
- (a) 
All samples with nitrogen concentration below 120~ppm  show unconventional 
magnetic moment behavior in the field hysteresis and temperature dependence  below $\sim 30$~K.
As example, we show in Fig.~\ref{ht} the field hysteresis of one of the measured samples in this study,
before and after annealing. Above $\sim 30$~K all samples behave as a typical undoped
diamagnetic diamond. 

- (b) The irreversible magnetization of the diamond samples 
can be phenomenologically understood as the
superposition of a diamagnetic, superparamagnetic (or ferromagnetic) and superconducting
contributions. We stress that the main characteristics of the anomalous hysteresis are also observed in ferromagnetic/superconducting oxide bilayers for fields applied parallel to the
main area (and interface between) of the two layers, see Fig.~\ref{ht}. 
More hysteresis
loops can be seen in   \cite{bardia} and in its supplementary information. The similarities in the minutest details between the
field loops obtained at different temperatures starting both in the virgin, zero field cooled states, are
striking. Note that we are comparing two data sets with very similar absolute magnetic moments using
the same SQUID magnetometer. The strength of the magnetic moment signal does indicate that we are dealing with a 
 huge effect, 5 orders of magnitude larger than the sensitivity of our SQUID magnetometer (of the order of
 $10^{-8}~$emu).
The similarities in the field hysteresis and also in the
temperature dependence of the magnetic susceptibility 
suggest the existence of   superparamagnetic and superconducting regions in the N-doped diamond crystals.\\ 

- (c) The amplitude of the anomalous magnetization signal below  $\sim 30$~K  increases with  the
nitrogen-related C-centres concentration. \\ 

- (d)
The obtained phase diagrams for both phases found in the N-doped diamond samples below  30~K indicate
that the ordering phenomena should have a common origin. In contrast to the amplitude of the anomalous magnetization, the
critical temperature - below which the hysteretic behavior is measured - does not depend significantly on the C-centre concentration.
This result suggests that   localized C-centre clusters with similar C-centre concentration trigger the anomalies below 30~K. \\ 
- (e) 
The magnetic and superconducting regions are of granular nature embedded in the  dielectric diamond matrix.
The  used  Bean-like
model  to fit the field hysteresis suggests the existence of superconducting shielding currents
and field gradients within the grains of size larger than a few nanometers. 
- (f) The  total mass of the
regions of the samples that originates the anomalous magnetization behavior was estimated as $\sim 10^{-4}$ of  total
sample mass, emphasizing its granular nature and the difficulties one has to localize them in large bulk samples. \\

 We note that nitrogen, as a donor impurity in diamond, is expected
to  trigger SC with higher critical temperatures than with  boron \cite{bas08}. This prediction is based on the  higher binding energy
of substitutional nitrogen  in diamond, in comparison to boron. However, if ones takes boron as example, a nitrogen concentration of 
a few percent appears
to be experimentally difficult to achieve in equilibrium conditions. Therefore, it has been argued that disorder or defective regions and an inhomogeneous  N-concentration
in the diamond lattice may play a  role in triggering granular  SC \cite{bas08}.

Interestingly, compression-shear deformed diamond was recently
proposed as a new way to trigger phonon-mediated SC up to 12~K  in undoped diamond \cite{liu20}. 
Defect-induced superconductivity is actually  not a new phenomenon but was already reported in different
compounds. For example, 
strain-induced SC at interfaces of semiconducting
layers has  been treated theoretically
based on the influence of partial flat-bands \cite{tan014}.
SC was found in semiconducting superlattices up to 6~K and  attributed to
dislocations \cite{fog01} at the interfaces \cite{fog06}. 
Furthermore, SC was discovered at artificially produced interfaces in Bi and BiSb bicrystals having a 
$T_c \lesssim 21~$K, whereas neither Bi nor BiSb are superconducting \cite{mun08}. 
Finally, SC at two-dimensional interfaces  
has been recently  studied at the 
stacking faults of pure graphite or multilayer graphene structures theoretically \cite{kop13,mun13,pam17}
and experimentally \cite{bal13,cao18}.

\subsection{Aims of the present work}

Because of the
large penetration depth at the selected electron energy, we use a weak electron irradiation to study the influence of 
lattice defects produced or changed by the irradiation 
on the magnetization of  bulk N-doped diamond
crystals. Weak electron irradiation means that its fluence is
of the order of the concentration of C-centres, i.e., a few tens of ppm. 
Magnetization measurements  
can provide valuable information especially      in the case the phase(s) at the origin of the anomalies of interest, is (are) not homogeneously 
distributed in the samples.  
One  aim of this work is therefore to check with temperature and field dependent measurements of the magnetization, whether its anomalous 
behavior found in N-doped diamond crystals
can be affected by a weak electron irradiation.

If the irradiation does change the anomalies observed in the magnetization, which lattice 
defect  is mostly affected by the irradiation?  Can it be correlated to the magnetic anomalies?
A recently published study  on the influence of ion irradiation on the SC of heavily B-doped diamond samples, showed 
that SC is suppressed  after 
He-ion irradiation of  $5 \times 10^{16}$ at 1-MeV producing
 a vacancy concentration of $\simeq 3 \times 10^{21}$cm$^{-3}$ ($\sim 2\%$) \cite{cre18}. 
This apparent complete suppression of SC after irradiation was experimentally observed 
by electrical resistance measurements. Therefore, we may further ask, if the 
nominal concentration of produced vacancies by the irradiation that affects the magnetic properties 
is also of the order of  the C-centres concentration.
Furthermore, can a  high-temperature annealing  in vacuum of the irradiated samples recover  
  the anomalous behavior in the  magnetization?

\section{Experimental details}
\subsection{Samples and characterization methods}

We  selected four (111) diamond bulk  crystals, three of them were previously studied in detail  \cite{bardia}. 
The N-doped diamond single crystals were obtained  from the Japanese company SUMITOMO.
The crystals
were cleaned with acid (boiling mixture of nitric acid (100\%), sulphuric acid (100\%) and perchloricacid (70\%) 
with a mixing ratio of 1:3:1 for 4~h). All four ``as-received" samples were cleaned before any measurement
was started. The 
 presence of magnetic impurities was characterized by 
particle induced X-ray emission (PIXE) with ppm resolution in a diamond sample with similar magnetic characteristics
as the samples we used in this study  \cite{bardia}. The PIXE results indicated an overall  impurity concentration
below 10~ppm with a boron concentration below the resolution limit of 1~ppm.  
Table~I shows further characteristics of the samples.

  Because of the dielectric properties of the diamond 
matrix we are actually rather restricted to measure the magnetic moment of the samples to
characterize their magnetic response as a function of field and temperature. 
Even after irradiation the samples remain highly insulating, although their
yellow color gets opaque.  The magnetic moment of
the samples was measured using a superconducting quantum interference
device magnetometer (SQUID) from Quantum Design. Due to the expected granularity of the phases responsible for
the anomalous behavior of the magnetization, local magnetic measurements (MFM or micro-Hall sensors, as example)
are certainly of interest and in principle possible. However, assuming that the phases of interest are at the near surface region, 
the relatively large area of the samples (some mm$^2$, see Table~I) 
makes local measurements  time consuming, typically several months for an area of 1~mm$^2$ and at a given temperature and
magnetic field.

As in \cite{bardia}, the presence of regions with internal stress was investigated by polarized light microscopy. 
Within experimental resolution those
regions were not significantly changed after electron irradiation. 

\begin{widetext}

\begin{table}
 \caption{Measured samples with their main characteristics and treatments. ($\star)$: These samples were previously characterized in \cite{bardia}. 
      The concentration of paramagnetic (PM)-centres from SQUID measurements  was obtained from the linear slope 
    of the magnetic susceptibility vs. $1/T$ assuming $J = 1/2$ and $g_J = 2$
    and has an experimental error  $\lesssim \pm 7~$ppm. 
    The error  of  the average C-centres concentration obtained from EPR is   
    $\sim 15\%$.
    The written range of the C-centres concentrations (error $\lesssim 5\%$) 
    obtained from IR-absorption spectroscopy  represents the minimum and maximum concentrations  
    measured at different locations of the same
    sample, an indication of the inhomogeneous N-distribution. 
         The electron irradiation was performed in vacuum
        and at $900^\circ$C. Each annealing treatment 
        was  4~h in vacuum at 1200$^\circ$C with a heating and cooling time of 10~h. }
        \begin{tabular}{@{}llllllll@{}}
    \hline
 &Name & Size & Mass & PM & C & C & Treatment \\
  &&&&SQUID&EPR&IR&\\
      &&mm$^3$ & mg &  ppm  & ppm & ppm &\\
    \hline
 1& CD2318-02 & $2.4 \times 2.4 \times 1.7$ & 34.1 &- & 40&$33 \ldots 47$ &as-received\\
      \hline
  2& CD2318-01$^\star$ & $2.4 \times 2.4 \times 1.7$ & 34.0&- &-&$60 \ldots 65$& {as-received} \\
 &&   & &80 & 20 & $15 \ldots 27$&e$^-$-irrad.~($2\times10^{18}$cm$^{-2}$)  \\ 
 &&   &&76 & -& $13 \ldots 25$&annealed\\
       \hline
     3&  CD1512-02$^\star$ & $1.6 \times 1.5 \times 1.2$ & 9.8 & - & -&$18 \ldots 25$& {as-received} \\
  && & & 40&- &-&  e$^-$-irrad.~($1\times10^{18}$cm$^{-2}$)  \\
  &  & & &39&10&5 \ldots 10&annealed \\
  & & &&44& -&- & annealed \\
   \hline
   4& CD2520$^\star$ & $2.5 \times 2.5 \times 2$ & 44.1&-  &-&26 \ldots 40& {as-received} \\
  && &&- &-& -&  annealed \\   
    \hline
  \end{tabular}
   \label{table1}
\end{table}

\end{widetext}

Continuous wave (cw) EPR measurements were performed at room temperature with a BRUKER EMX Micro X-band spectrometer  at 9.416$\,$GHz using a BRUKER ER 4119HS cylindrical cavity.  Absolute concentrations of paramagnetic C-centres (usually called P1-centres in EPR) and NV$^-$-centres were determined in dark by using an ultramarine standard sample with known spin number and double integration of the corresponding EPR spectra. The external magnetic field $B_0$ was oriented along the [111] crystallographic axis of the diamond single crystals. In order to avoid saturation effects a 10 kHz modulation frequency at a microwave (MW) power of  $P_{MW}$ = 630 nW was employed. We verified that the EPR signal intensities $I_{EPR}$ of both centres deviate from a square root dependence on  $P_{MW}$  by less than 10 \% between  $P_{MW}$ = 203 nW and 630 nW and therefore saturation effects can be  neglected at room temperature and such low mw power levels. Modulation amplitudes of 0.1 mT and 0.03 mT were employed to
measure  the C- and  NV$^-$centres to avoid broadening effects.\\

 A reliable method to characterize the granular, inhomogeneous concentration of nitrogen and the N-based centres, like the  C-centres (neutral nitrogen N$^0$ with a maximum absorption at 1344~cm$^{-1}$) and N$^{+}$ (positively charged single substitutional nitrogen with a maximum absorption at 1332~cm$^{-1}$), is infrared (IR) microscopy. IR measurements and IR spectral imaging were carried with a Hyperion 3000~IR microscope coupled to a Tensor~II FTIR spectrometer (Bruker Optik GmbH, Ettlingen, Germany). The microscope is equipped with both, a single element MCT detector and a $64 \times 64$ pixel focal plane array (FPA) detector. Measurements with the MCT detector were carried out with a spectral resolution of 1~cm$^{-1}$, whereas the resolution of the FPA detector used for imaging was limited to 4~cm$^{-1}$. Diamonds were fixed in a Micro Vice sample holder (S.T. Japan) and carefully aligned horizontally. 

The concentration values of  C-centres shown in Table~I were obtained from IR transmission spectra taken with the MCT detector at various positions of the diamond surface. The MCT detector was favored over the FPA detector for these measurements due to the very low band widths of the above-mentioned bands. Quantitative concentration data of N$^+$- and C-centres were derived from the spectra using  the relationships between the peak intensities of these bands and the concentration of corresponding nitrogen centres given by Lawson et al. \cite{law98}, see Section~\ref{is}.

\subsection{Electron irradiation}
\label{ei}

Electron irradiation was done at 10~MeV energy using a B10-30MPx Mevex Corp. (Stittsville, Canada) with a total irradiation fluence 
 of $1 \times 10^{18}$~electrons/cm$^2$ (sample 3) and 
$2 \times 10^{18}$~electrons/cm$^2$ (sample 2). The electron irradiation was performed   at 900$^\circ$C in vacuum, in order to remove a certain amount of
disorder. 
In spite of a large number of studies on irradiation effects in diamond, the available literature on
the electron irradiation damage   in diamond is  less extensive. 
For example, electron irradiation at high temperatures was used
to significantly increase  the density of nitrogen-carbon vacancy (NV) centres \cite{aco09,cap19}. 
Several characteristics  of the electron irradiation damage  in diamond 
were published by Campbell and Mainwood \cite{cam00}. From that study, we estimate a
  maximum penetration  depth of $\simeq 15~$mm  at the used electron energy. 
  This indicates that the irradiated electrons completely penetrate the used samples.\\

 There are two recent works directly related to our studies. One of them investigated
the effects of electron irradiation on in-situ heated nano-particles of diamond \cite{min20}. These samples were 
 irradiated with the same accelerator as our crystals. 
In \cite{cap19} the authors increased the NV-centres concentration through electron irradiation maintaining the bulk diamonds 
at high temperatures. Both studies showed a higher conversion rate of C-centres into NVs with in-situ annealing. 
Taking into account  those studies, we  used  10~MeV electrons irradiation to  convert C-centres into NVs. 
The selected  fluences  were chosen to produce a {\em nominal} concentration of vacancies similar to the  
concentration of magnetic C-centres existent before irradiation.  
In this way, we can directly check, whether a change in the concentration of C-centres affects or not, the magnetization of our samples.
  
In terms of created vacancies, homogeneously induced by the electron irradiation, the 
maximal nominal concentration would be  $5 \times 10^{18}~$vac./cm$^{3}$ for sample 3, i.e.  a concentration of 
$\simeq 30$~ppm (60~ppm for  sample 2),  of the same order as the
 C-centres obtained by IR and EPR, see Table~I. 
Obviously, the  vacancy concentration that remains in the samples is smaller because of 
the high temperature of the samples  during the irradiation
process \cite{new02}. It means that some of the vacancies 
can give rise to N-related defects and  others  diffuse to the sample surface. 
From the change in the concentration of NV-centres one can estimate the vacancy concentration
that remains.

\section{Electron Paramagnetic Resonance}
\label{epr}
\begin{figure}
	\includegraphics[width=1\columnwidth]{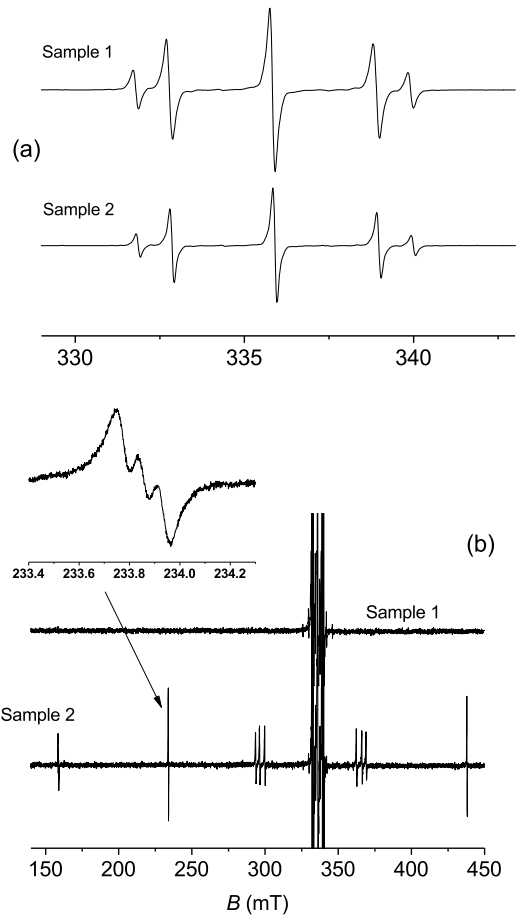}
	\caption{EPR spectra obtained at room temperature of samples~1 (as-received)  and 2 (after irradiation), see Table I, 
	for  $\bf{B_0} \|$[111] showing the signals from: (a) C-centres and (b) from NV$^-$-centres. 
	The spectrum of the NV$^-$-centres were recorded with a tenfold higher receiver gain. 
	The insert in (b) displays the $^{14}$N hyperfine splitting of the $m_S  = +1 \leftrightarrow 0$  signal of the NV$^-$-centres at 233.8 mT.}
	\label{fig:epr1}
\end{figure}

Figure~\ref{fig:epr1} illustrates EPR spectra of the diamond single crystal samples 1 and 2 recorded for $\bf{B_0} \|$[111]. Both samples show the intense signals of the C-donor of nitrogen incorporated at carbon lattice sites having an electron spin $S = 1/2$ and the typical  $^{14}$N hyperfine (hf) splitting into three lines  
(Fig.\ref{fig:epr1}(a)) \cite{lou78}. For the chosen orientation of the single crystals the external magnetic field $\bf{B_0}$ points along the $C_3$ symmetry axis of one of the four magnetically nonequivalent C sites in the diamond lattice \cite{lan91}. The $C_3$ symmetry axis of the C-centres defines the symmetry axes of its axially symmetric $g$ and $^{14}$N hf coupling tensor. The two outer hf lines at 331.8 and 339.9 mT together with the central hf line at 335.8 mT belong to this orientation of the C-centres. The symmetry axes of the other three sites of the C-centres make an angle of 109.47$^\circ$ with $\bf{B_0}$ and lead to hf lines at 332.8~mT and 338.9~mT in addition to the central hf line at  335.8~mT. Sample 3  showed a comparable spectrum of the C-centres. The  concentrations  of the C-centres in the three diamond single crystal samples 1, 2, and 3  (Table I) were determined by double integration of the full EPR  spectrum taking into account the quality factor of the cavity and comparison with an ultramarine standard sample with known spin number.
We emphasize that the concentration of C-centres obtained with this method is similar to the one obtained by
IR spectroscopy, see Table I and  next section.

A broader magnetic field scan with a larger amplification as done for  sample 2  (Fig.\ref{fig:epr1}(b)) revealed 
additional signals of the NV$^-$-centres with $S = 1$ \cite{lou78}.  Note that the intense signals at about 336~mT in Figure~\ref{fig:epr1}(b) are due to C-centres. The NV$^-$-centres has been assigned to a nitrogen atom substituting a carbon lattice site, which is associated with a next neighboured  carbon vacancy.  NV$^-$-centres could not be observed for  sample 1. The $C_3$ symmetry axis of the  
NV$^-$-centres is oriented along the [111] direction and determines the symmetry
 axis of the axially symmetric zero field splitting tensor of this centres. In  samples 2 and 3, the concentration of NV$^-$-centres 
 is  $\sim(2 \pm 0.3)$~ppm. Therefore, we assume that these centres
 do not play any role in the phenomena we discuss in this work.

\section{Infrared spectroscopy}
\label{is}
\begin{figure}
	\includegraphics[width=1\columnwidth]{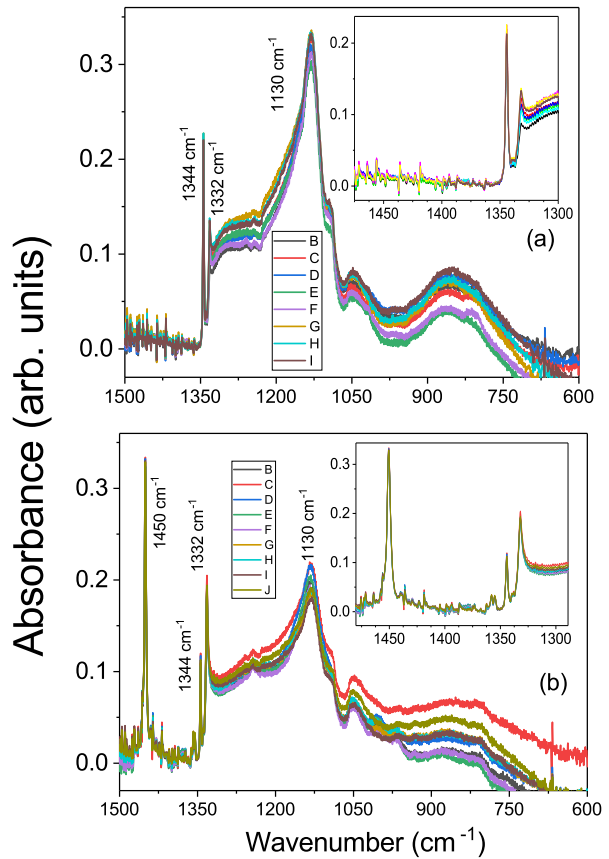}
	\caption{(a) Infrared spectra of the absorbance vs. wavenumber of sample 1 in the as-received state measured in transmission with a
	resolution of $1$~cm$^{-1}$.  The inset shows a blow out at the high wavenumber region. (b) Similar for sample 2 after electron irradiation. The different curves (B $\ldots$ J) in both figures were obtained at different positions of the sample. } 
	\label{ir}
\end{figure}

In a first published study of similar diamond crystals, a correlation between the magnitude of 
the anomalous magnetization below 30~K and the C-centres was found \cite{bardia}. 
Therefore, we  characterized the concentration and the space distribution of  this magnetic defect.  
The characterization of this centres in diamond using IR spectroscopy  has been reported in several earlier studies \cite{boy95,law98,hai12,kaz16}, including radiation damage and subsequent annealing \cite{col09}.
The C-centres are responsible for the broad absorption peak at 1130~cm$^{-1}$ 
followed by a very sharp absorption maximum at 1344~cm$^{-1}$, see Fig.~\ref{ir}. The broad one at 1130~cm$^{-1}$ is attributed to a quasi-local 
vibration at single substitutional nitrogen atoms \cite{bri91,law98,hai12,kaz16}, whereas the one at 1344~cm$^{-1}$  is due to a local mode of vibration of the carbon atom 
at the C-N bond with the unpaired electron \cite{col82}.

 \begin{figure}
	\includegraphics[width=1\columnwidth]{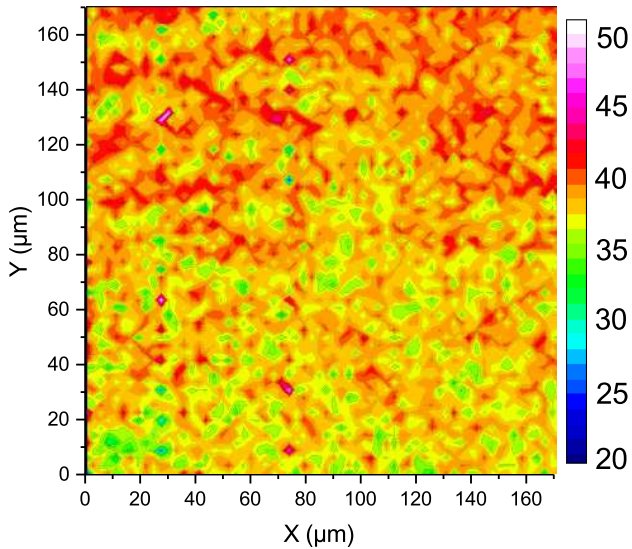}
	\caption{Distribution of C-centres in an area of $(170 \times 170)~\mu$m$^2$ of sample~1, 
	obtained from the maximum intensity of the IR band at 1344~cm$^{-1}$ using a FPA detector with
	a resolution of 4~cm$^{-1}$. The color  scale indicates the concentration of C-centres in ppm. 
	Note that this 
	two-dimensional distribution is the sum of the whole absorption through the whole
	sample thickness at the selected
	energy. The concentration values in the color scale are in ppm.} 
	\label{CI}
\end{figure}

\begin{figure}
	\includegraphics[width=1\columnwidth]{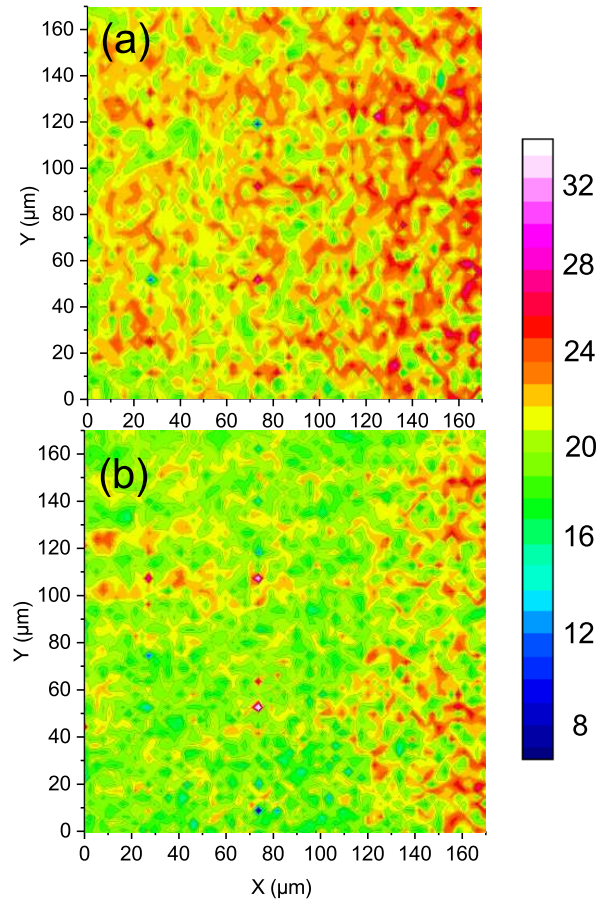}
	\caption{Distribution of C-centres in an area of $(170 \times 170)~\mu$m$^2$ of sample~2: (a) after electron irradiation and (b) after annealing measured at the same location  using a FPA detector with
	a resolution of 4~cm$^{-1}$. Data were obtained from the maximum intensity of the IR band at 1344~cm$^{-1}$. 
	The color  scale indicates the concentration of C-centres in ppm.} 
	\label{CI-2}
\end{figure}

The concentration range of C-centres determined from IR spectra and measured at different positions 
of the samples is given in Table~I. The values were obtained using the relationship  C(ppm) $= 
37.5~A_{1344\text{cm}^{-1}}$ \cite{law98}, where $A_{1344\text{cm}^{-1}}$ is the peak intensity 
 at the corresponding wavenumber. The concentration range of N$^+$-centres 
can be estimated  from the absorption peak at 1332~cm$^{-1}$ (using N$^+$ (ppm) = 5.5~A$_{1332\text{cm}^{-1}}$) \cite{law98}. 
In all samples and states of the samples, the N$^+$ concentration is much smaller (factor 3 to 10) than the C-centres concentration and
therefore it will not be taken into account in the  discussion of the results. 
 We note that the range of  C-centres concentration obtained from
 IR-absorption  agrees very well with the average concentration obtained from EPR, see Table~I. 

As example, we show in Fig.~\ref{ir} the IR spectra of: (a) sample 1 in the as-received state and (b) sample 2 after electron irradiation. 
All spectra were obtained in transmission (MCT detector) with a resolution of $1$~cm$^{-1}$ at different positions of the sample surface.
 After electron irradiation,  a new maximum at 1450~cm$^{-1}$ appears, see Fig~\ref{ir}(b). 
We note that 
in the as-received state of all  samples this maximum is absent within experimental resolution, see Fig.~\ref{ir}(a) as example.  

The maximum absorption at 1450~cm$^{-1}$ was already shown to 
develop during the annealing at $T > 500^\circ$C in  type I diamonds after  electron-irradiation and is related to N-interstitials 
in the diamond lattice \cite{woo82}. Since then, it has been studied and reported several times \cite{kif96,gos04,bab16}. 
Last published characterization measurements indicate that the origin of this maximum is related to 
 H1a-centres (di-nitrogen interstitials) \cite{gos04,bab16}. The vanishing or grow of H1a-centres appears to be
correlated to the interplay between C- and N$^+$-centres aggregation processes. The results in  \cite{bab16}
suggest that an increase in the density of H1a-centres is accompanied by a decrease in the density  of C-centres. 

In order to quantitatively estimate the extend of the increase of the concentration of H1a centres [H1a], we use 
a relation given by Dale \cite{dal15}, which relates the band integral  $A$ at 1450~cm$^{-1}$  to the concentration of interstitials
according to [H1a](ppm) $ = 
		A$ (in cm$^{-1}) / f$,  
with $f = 3.4 \times 10^{-17}~$cm. Spectra taken with the MCT detectors at various positions of the irradiated diamond before and after annealing were baseline-corrected and normalized according to \cite{dal15} before conversion of the band integrals to [H1a]. Finally, the individual values of [H1a] were averaged. Using this procedure for the results of sample~2, we found a slight increase of [H1a] from $\simeq 3$~ppm before annealing to $\simeq 3.3$~ppm after annealing. The  values  indicate that
   [H1a~]$\lesssim 15\%$ of the C-centres concentration.

Because the concentration of C-centres plays a main role in the origin of the phenomena observed in the magnetization, 
it is of interest to display the concentration distribution of these centres. Figure~\ref{CI} shows one example of this distribution
measured in sample 1 (as-received state) recorded in transmission with the FPA detector, i.e. the image accounts for the absorption around 1344~cm$^{-1}$ measured through the whole  sample thickness with a spectral resolution of  4~cm$^{-1}$. 
This image shows a remarkable  granular-like distribution of the concentration in the selected area 
of $170~\mu$m~$ \times 170~\mu$m. This image   indicates the existence of regions of a  few $\mu$m$^2$ with clear differences in the concentration
of C-centres of up to $\sim 25\%$ between some neighbouring regions, see Fig.~\ref{CI}.  According to the IR absorption peak at 1344~cm$^{-1}$, the concentration
of C-centres can reach values up to  $\sim 50~$ppm locally, depending on the region of the sample.

\begin{figure}
	\includegraphics[width=1\columnwidth]{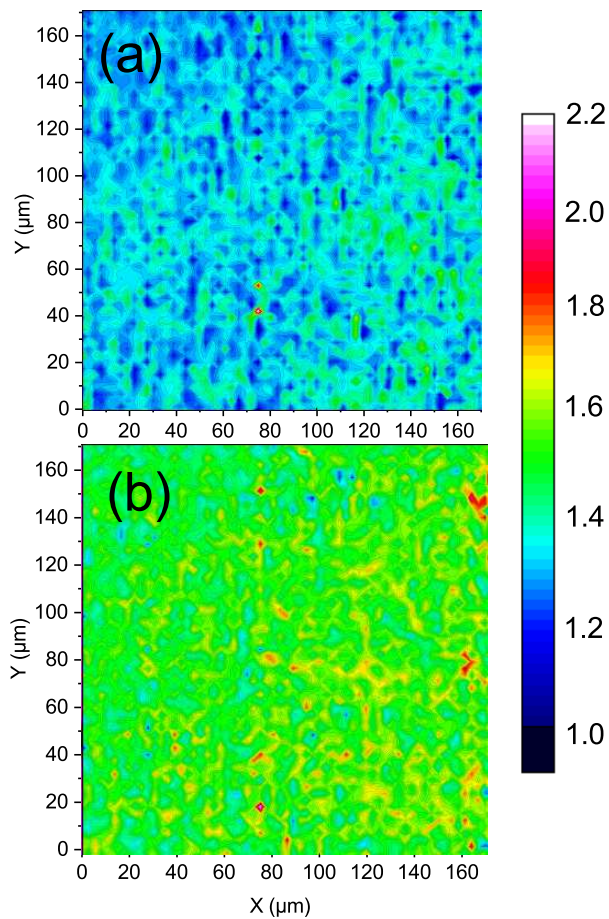}
	\caption{Similar to Figs.~\ref{CI} and \ref{CI-2} but for the IR absorption at 1450~cm$^{-1}$  due to H1a-centres of sample~2: (a) after electron irradiation and (b) after annealing. The color  scale indicates the amplitude  of the absorption maximum at 1450~cm$^{-1}$. To approximately estimate the concentration in ppm, multiply the numbers at the color scale by two.} 
	\label{CI-3}
\end{figure}

Figure \ref{CI-2} shows the concentration distribution of C-centres of sample~2 after electron irradiation (a) and subsequent annealing (b), see also Table~I,
both measured at the same  sample area. The distribution of  the H1a-centres correlated to the maximum at 1450~cm$^{-1}$, after electron irradiation (a) and
after subsequent annealing (b) at the same sample area,  is shown in Fig.~\ref{CI-3}. 

Several details are of interest, namely: (1) The average concentration of C-centres  after electron irradiation decreases by $\simeq 50\%$ and shows a similar granular distribution as in the as-received samples but with a   larger  difference (up to 50\%) between neighboring regions. 
(2)  After annealing, see  Fig.~\ref{CI-2}(b),   the
average concentration of C-centres decreases further by $\sim 6\%$ but shows a more homogeneous
distribution, see also Table I.  (3) The concentration of H1a-centres (see Fig.~\ref{CI-3}), which appear only after irradiation and annealing \cite{woo82},  
becomes homogeneously distributed and  its average density 
 increases by $\sim 10\%$  after annealing.

\section{Magnetization results}
\subsection{Field hysteresis loops}

\begin{figure}
\includegraphics[width=1\columnwidth]{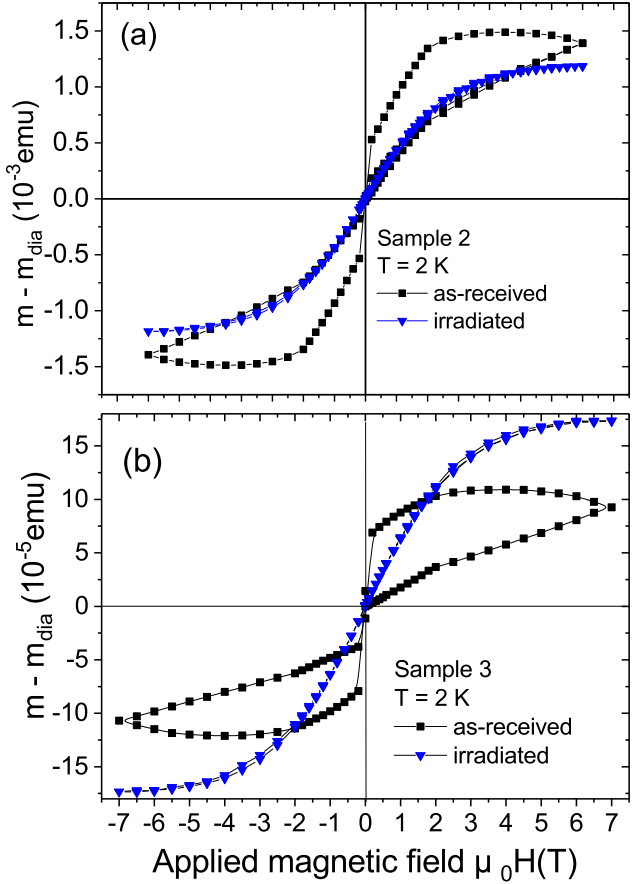}
\caption{\label{2K-mH} Field hysteresis loops at $T = 2~$K of the  
magnetic moment of samples (a) 2 and (b) 3, before and after irradiation. Note the small remaining hysteresis between 2~T and 4~T in the irradiated sample 3. 
The diamagnetic background obtained at 300~K was subtracted from the measured data. The measurements of  sample 2 were done in two
field directions with similar results within resolution. The direction of the applied field shown in all figures is $\perp $ [111]. }
\end{figure}

The field hysteresis loops were measured at constant temperature after zero field cooling from 380~K.
The
field was swept from $0~\rm T\rightarrow +7$~T$\rightarrow -7$~T$\rightarrow $+7~T.  After this
procedure, the applied field was set to zero, the sample heated
to $380$~K  and cooled at zero field to
a new  constant temperature.
The main characteristics of the anomalous hysteresis at 2~K can be seen  in Figs.~\ref{2K-mH} and \ref{ht}, 
observed in all samples before irradiation.  
The magnetic moment $m$ of the diamond samples shown in Figs.~\ref{2K-mH}  and \ref{ht}
was measured with the large
surface areas of the samples parallel  to the magnetic field applied $\perp$ [111]. Measurements
at other field directions were also done in order to check for the existence of any magnetic anisotropy.
The magnetic moment did not show any anisotropy within error, in agreement with the reported data in \cite{bardia}. 

We note that the high sample-temperature stabilized  during irradiation  is not the reason for the changes we discuss below. In order to prove this, we
show in Fig.~\ref{ht} the magnetization field hysteresis loop at 2~K of sample~4 before and after annealing at 1200$^\circ$C in vacuum for 4 hours.
The field loops  indicate that the main field hysteresis  does not change significantly after  annealing. 

The influence of the electron irradiation on the field hysteresis can be clearly seen in  Fig.~\ref{2K-mH}. The field 
hysteresis is suppressed. The  s-like field loop curve obtained for both samples at $T = 2~$K 
after irradiation is due exclusively to a paramagnetic (PM) phase, not SPM, as the measured temperature 
dependence of the magnetic moment also indicates, see section~\ref{td} below.

\begin{figure}
\includegraphics[width=1\columnwidth]{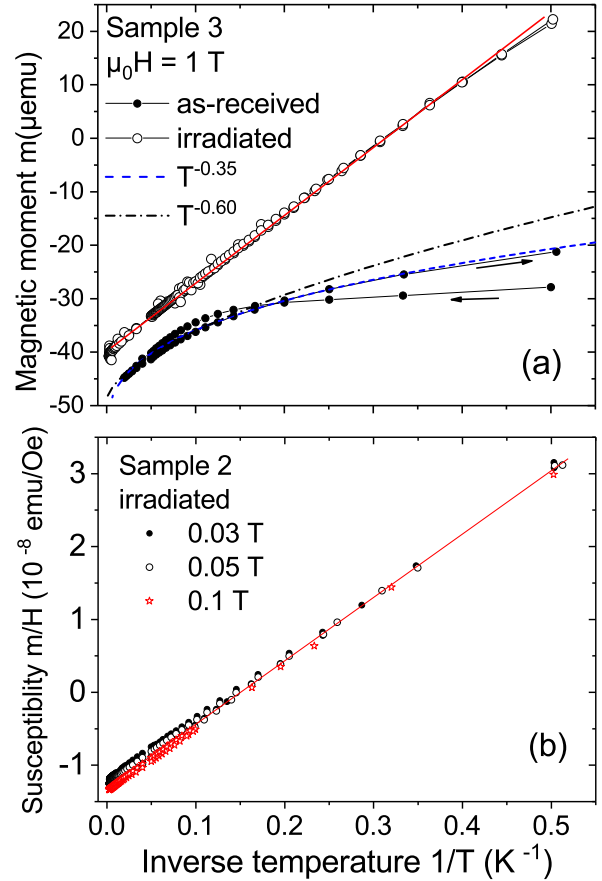}
\caption{\label{t-dep} (a) Magnetic moment vs. inverse temperature measured at a constant  field of 1~T 
for sample 3 in the as-received  and irradiated states. Note the temperature hysteresis between the zero field cooling (ZFC) (increasing temperature path) 
and the field cooling (FC) (decreasing temperature path) states. The dashed  line follows $m(T) = 45 T^{-0.35} - 56$ $[\mu$emu] 
and the dashed-dotted line  $m(T) = 52 T^{-0.6} - 49$ $[\mu$emu] with $T$ in K.  (b) Magnetic susceptibility,  defined as the ratio between the magnetic moment and the
applied field,  vs. inverse temperature for sample 2 in the irradiated state at three magnetic fields. No background was subtracted from
the raw data. The straight line is a linear fit to the data. }
\end{figure}

\subsection{Temperature hysteresis loops}\label{td}

To  further demonstrate that the s-like field loop observed after irradiation, see Fig.~\ref{2K-mH}, is related to a  PM  phase, 
we discuss  the measured
temperature dependence before and after irradiation. Figure~\ref{t-dep}(a) shows the magnetic moment of sample 3 before and after irradiation
vs. inverse temperature obtained at an applied field of 1~T. As discussed in detail in \cite{bardia}, the Curie-like behavior observed at relatively high temperatures, 
i.e. at $T > 20~$K, see Fig.~\ref{t-dep}(a), does not scale
with the applied field as one expects for a PM phase. Its susceptibility clearly increases with the applied field  \cite{bardia}, indicating the development of 
a SPM phase at low temperatures. This SPM phase does not show any hysteresis in field in the measured temperature range, although its s-shape in the field hysteresis loop looks
 similar to a FM state. The clear hysteresis  in the temperature loop,
see Fig.~\ref{t-dep}(a), has a different origin. In \cite{bardia} it has been interpreted  as due to a SC contribution added to the SPM one. 

After electron irradiation we observe remarkable changes in the temperature dependence of the magnetic moment, namely:\\ 
(a) The main change after irradiation 
is observed at   $T < 30~$K, where 
the large hysteretic  contribution  is completely   suppressed.
The magnetic moment shows a Curie-law behavior  in 
the whole temperature range, with a weak tendency to saturation at the lowest measured temperatures at an  applied field of 1~T,  see Fig.~\ref{t-dep}(a), 
as expected for a PM phase.\\ 
(b) From the slope of $m$ vs. $1/T$ and assuming
a total angular moment $J = 1/2$ and $g_J = 2$ per PM centre, we estimate a concentration of  $(40\pm 5)$~ppm PM-centres 
 for  sample~3 after irradiation with a fluence  of $1 \times 10^{18}$~cm$^{-2}$. 
This value can be interpreted as due to the following contributions: one from the C-centres with 
a concentration 
of   $\sim 10$~ppm, see Table~I, H1a-centres   
of   $\lesssim 5$~ppm concentration  and a rest  of $\sim 25$~ppm of magnetic centres due either vacancies and/or lattice defects produced by them. \\
(c) The results obtained from sample 2 after irradiation are shown in  Fig.~\ref{t-dep}(b) where the susceptibility vs. inverse temperature
at three applied fields is plotted. The temperature dependence and the observed scaling with applied field are  compatible with a  PM phase, see Fig.~\ref{t-dep}(b).  
From the susceptibility
slope vs. $1/T$ of sample~2 and after two times higher  electron irradiation fluence  the 
calculated density of PM centres  is $(80 \pm 5)$~ppm. It means 
a factor of four larger than the density of C-centres obtained by EPR and IR, see Table~I. This indicates that about 60~ppm 
of the  PM-centres measured by the
magnetic susceptibility should be  related to extra lattice defects induced by the irradiation, 
with a magnetic moment of the order of $1 ~\mu_B$.\\
 (d) After an electron irradiation with a fluence that 
produced a  decrease of a factor of two  in the C-centre concentration (equivalent to some tens of ppm), 
any anomalous signs of hysteresis in field and temperature are completely absent 
 at $T \ge 2~$K, within experimental resolution. \\

\subsection{Partial recovery of the anomalies in the magnetization after high temperature annealing in vacuum}
\label{partial}

\begin{figure}
	\includegraphics[width=1.1\columnwidth]{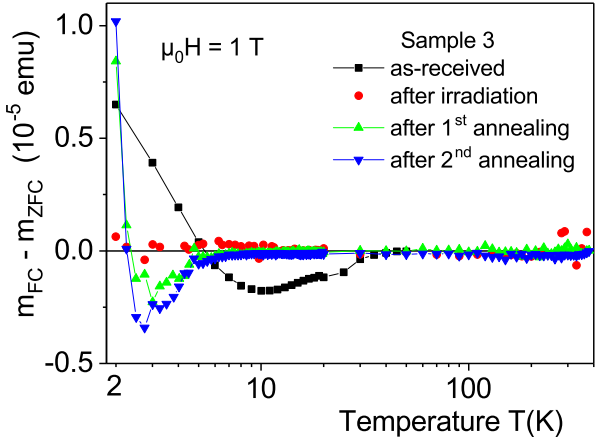}
	\caption{Temperature dependence of the  difference between the field cooling (FC) and zero field cooling (ZFC) states $m_d = m_{FC} - m_{ZFC}$ measured in sample~3 in the as-received, after irradiation and after
	the first and second annealing treatments at a constant applied field of 1~T. }
	\label{th}
\end{figure}

In an earlier study, Creedom et al. \cite{cre18} found that the observed suppression of SC 
of B-doped diamond  after 
a certain fluence of He-ion irradiation, could be partially recovered after annealing the sample in vacuum. Therefore, we  annealed 
some of the samples in vacuum at 1200$^{\circ}$C for 4~h with a 10 hours ramp up and down, following the sequence used in
\cite{cre18}.  Figure~\ref{th} shows the difference between the magnetic moment at FC and at ZFC states of sample 3
obtained at constant field of 1~T  in the as-received, after irradiation and after two identical annealing
processes. A finite, negative difference in $m_d = m_{FC} - m_{ZFC}$ is observed at 5~K~$\lesssim T \lesssim 35$~K in
the as-received state, in agreement with the earlier publication \cite{bardia}. An interpretation for this anomalous difference 
is given in the discussion.

 After electron irradiation $m_d =  (0\pm 1)~\mu$emu in the whole temperature range, see Fig.~\ref{th}.
 After the first annealing  a similar behavior as in the as-received state is observed but   $m_d$ starts 
 to be negative at 
 $T < 10~$K showing the minimum at $\sim 3~$K (instead of 10~K as in the as-received state). 
 A second annealing slightly increases  $|m_d|$ with a small shift to higher temperatures.  
 The obtained behavior of $m_d(T)$ after annealing indicates the partial recovering of the regions responsible for
 the anomalous behavior of the magnetization.  
 
\begin{figure}
	\includegraphics[width=1.15\columnwidth]{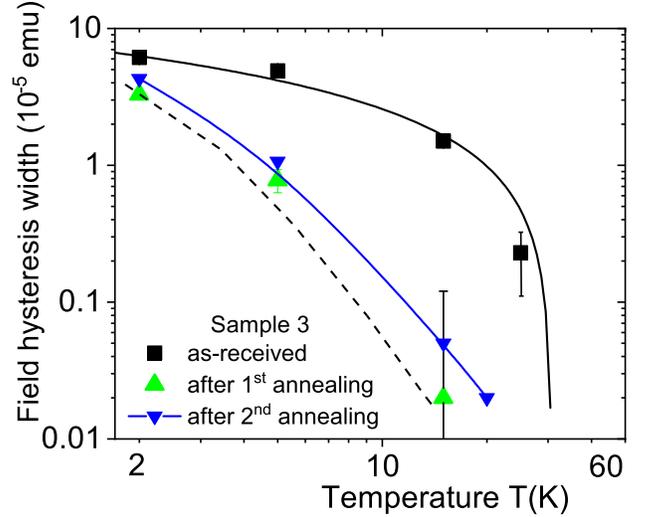}
	\caption{Field hysteresis width at a constant field of 3~T  measured at different temperatures for sample 3 in 
	the as-received, first  and second annealing treatment after electron irradiation. The black line follows the  
	equation\\ $0.05 - 2.3 \ln(T/T_c)$ with $T_c=30$~K. This  phenomenological equation was found to satisfactorily describe the critical line of  
	 several as-received N-doped diamonds \cite{bardia}. The blue and dashed lines through the data points of  the annealed 
	 samples are only a guide to the eye.}
	\label{fh}
\end{figure}

 This partial recovering  is also observed by
 measuring the field hysteresis loops after annealing, in particular the field hysteresis width. 
 Figure~\ref{fh} shows the  field hysteresis  width  at a constant field
 of 3~T vs. temperature for the as-received and after the two annealing treatments after irradiation of sample~3. The results indicate
 that the amplitude of the anomaly in the magnetization  partially recovers after annealing,  see 
 Fig.~\ref{fh}. However, whether a   critical temperature is really reduced respect to the one  in the as-received state is 
 not so clear because of the different temperature dependences, see 
 Fig.~\ref{fh}.  An apparent reduction of the superconducting $T_c$ obtained from transport measurements    has been reported 
 after subsequent annealing treatment of a previously He-ion-irradiated
 B-doped diamond sample \cite{cre18}.

\section{Discussion}

Pure diamond samples, without N- or B-doping, show no anomalous behavior in the magnetization
above 2~K but the usual diamagnetic response. 
A very weak Curie-like behavior in the magnetic susceptibility below 100~K is
observed in some of the ``pure" diamond samples, 
with a temperature irreversibility between the ZFC and FC states more than three orders of magnitude
smaller than the irreversibility we measured in N-doped diamond \cite{bardia}. 
Because the concentration of boron in all  diamond crystals presented in this work 
is below 1~ppm,  
we can certainly rule out B-doping as  responsible for the anomalies 
in the magnetization. 

A correlation between the anomalous maximum in magnetization and
the concentration of C-centres has been obtained in \cite{bardia}. 
Also our results indicate a  correlation to the
C-centres. But this correlation is not a simple one, as the vanishing and recovery of the anomalies and the 
corresponding C-centres concentration for those states indicate, see Table~I.


As emphasized in the introduction, the similarities between the magnetization loops obtained in N-doped diamonds and
FM/SC bilayers, see Fig.~\ref{ht}, suggest  that the anomalous behavior 
is related to the existence of both, a SC and a SPM phase, as proposed in the original publication. 
The way to simulate such a rather anomalous field hysteresis is given by the simple 
superposition of each field dependent magnetization plus eventually a small diamagnetic background, linear
in field. The good fits of the field hysteresis data of the diamond samples (as well as of the bilayers) 
to this model can be seen in  \cite{bardia}. Also the obtained dependence of  the fit parameters  in temperature 
supports the used model.

 At this point  we would like
 to give a couple of remarks on the EPR measurements. One would argue that below 30~K and if such a 
 mixture of SC and SPM phases  in our samples exists, then we would expect to see a broadening of the EPR line related to
 the C-centres at low temperatures. To check for this effect we have done 
 additional EPR experiments with samples 1, 2 and 3  at 11~K. 
A significant line broadening  for the signals of the C- and NV-centres  
was not be observed, in comparison with the room temperature spectra. 
The main reason for the absence of such a broadening of the EPR C-lines is the following.
EPR measures the average response of all C-centres, which are distributed through all the sample. 
From all those regions, only a small part of them have the phases responsible for the huge field (and temperature) hysteresis loops. 
With our EPR equipment, it is simply not possible to get the signal  of such a small amount. 
Note that there is always an overlap with the dominating signal from the C-centres located all over the remaining parts of the sample.

\subsection{How large would be a significant C-centres concentration to trigger ordering phenomena in  diamond ?}

This question is not answered experimentally in the literature. Surprisingly, with exception of \cite{bardia},
systematic bulk magnetization measurements of diamond samples
with different concentration of P1 or C-centres are not apparently published. Because the purity of diamond
crystals nowadays is high and especially magnetic impurities  can be quantitatively measured with high resolution (e.g.
with PIXE) there are no clear reasons why such measurements were not systematically done in the past.

 To estimate the average distance between C-centres 
we follow a similar procedure as  in \cite{hen19}, getting an average distance 
between C-centres  $<d>_{C-C} = (2.9 \pm 1.5)~$nm in the diamond matrix for a 
mean concentration of the order of 50~ppm. 
Because the concentration of C-centres is not homogeneously distributed in the diamond samples, it is
quite plausible that clusters  with an average smaller distance exist. 
Is such a distance 
between C-centres   in the diamond matrix too large to trigger ordering phenomena, even at low temperatures? 

Let us take a look at two recent publications that
studied the entanglement between single defect spins \cite{dol13} and NV$^-$-N$^+$ pair centre \cite{man18}.
In the first study deterministic entanglement of electron spins over a distance of 10~nm was demonstrated at room temperature. 
In the second work done in 1b diamond, the authors found that for all NV-centres with a neutral N within a distance of
5~nm, an electron can tunnel giving rise to NV$^-$-N$^+$ pairs. Taking into account these studies 
  a non-negligible coupling between C-centres with an average distance of less than 3~nm does not appear impossible.

A concentration of 50~ppm  of C-centres is much smaller than the necessary boron concentration ($\gtrsim 2~\%$) reported in the 
literature to trigger the observed 
superconducting transition in the electrical resistance. As we noted in the introduction,  new experimental studies
do suggest that the relationship between carrier concentration due to B-doping and the superconducting
critical temperature is not as clear. 
It may well be that the relatively high carrier concentration in those B-doped diamonds is necessary
to get  percolation, i.e. a superconducting current path between voltage electrodes. But  it does not necessarily  rule out that localized superconducting grains are
 formed at lower B-concentration.

\subsection{Granularity}

We may ask now, why the superconducting transition is not observed in the electrical resistance in our N-doped diamond samples  \cite{bardia}?
The reason is the granularity of the C-centres distribution;  it prevents the percolation of
a superconducting path between voltage electrodes when their distance is much larger than the grain size, see Fig.~\ref{CI}. 
This granularity  added to the apparently small 
amount  of the total sample mass that shows a superconducting response is the  reason why 
transport measurements are so difficult to perform in these samples.

From the comparison between the  absolute magnetic signals in the diamond samples and those  measured in oxides bilayers, a rough estimate of 
the equivalent mass of the SC phase in the N-doped diamond samples 
 gives  $\sim 1~\mu$g,
i.e.  the volume of a cube of  $70~\mu$m side length  \cite{bardia}. 
This estimate suggests that micrometer large
 clusters of superconducting C-centres 
should exist in the sample, as the image in Fig.~\ref{CI} suggests. 
 Note that the assumption  that all the measured anomalous magnetic signals would come from
very few,  well localized regions with a very high C-centres concentration in the percent region, as in the case of 
B-doped diamond, does 
not find support from our IR-absorption images. Also the clear suppression of 
the two phases after such a weak and homogeneous electron  irradiation and the thereof decrease  
of the C-centres concentration, speaks against such an assumption. The obtained evidence speaks for a correlation of 
the observed phenomena with localized clusters of C-centres with concentration at least ten times smaller than the
reported  B-concentration  in  B-doped diamond, see Section~\ref{intro} and references therein.

 We note that several experimental details gained from our magnetization results 
suggest that the SPM phase should emerge simultaneously to  the SC one. 
For example, the anomalous behavior of the difference $m_d = m_{FC} - m_{ZFC}$, see Fig.\ref{th}.  We note first that in the
case we would have only a superconducting single phase,  we  expect $m_d > 0$ at $T < T_c$. An interpretation for the anomalous behavior
of $m_d(T)$  has been provided in \cite{bardia} taking into account that
the samples have both,  a SC and a SPM phase. In this case the anomalously large increase of the magnetic moment
$m_{ZFC}$ is due to the  partial field expulsion in the  SC  regions.
In other words,   an effective higher field than the applied one enhances the magnetic moment of the SPM phase 
in the ZFC state and therefore  $m_d < 0$. At low enough temperatures, $m_{FC}$  increases and eventually $m_d > 0$, as observed.  
We emphasize that similar results were obtained in FM/SC oxide bilayers, supporting this
 interpretation \cite{bardia}.

\subsection{Possible origin for the existence of two antagonist phases within regions with high C-centres concentration}

Clustering of C-centres could produce a spin glass phase arising from their spin 1/2 character of independent 
magnetic moments. Following the ideas of the RVB model \cite{bas08,and87}, as temperature is reduced  the C-centres start to be weakly 
coupled and could form pairwise,  antiferromagnetically (AFM) 
 coupled singlets. A certain density of these donors may delocalize, leading to a SC state at low enough temperatures. 
   
There are at least two possible scenarios that could provide an answer to the simultaneous 
existence of the two phases. The rather conventional
possibility is that  at low enough temperatures, in our case $T \lesssim 30~$K, a mixture of
paired and  unpaired C-centres are formed, a  spin liquid. The unpaired C-centres start to 
contribute to the magnetic response under an applied magnetic field as a 
SPM phase, the reason for the s-like field loop contribution without any field hysteresis. 
A SPM phase, instead of a FM one, is possible only if the spin-spin interaction (inversely proportional to the
distance between the spins) is weak enough within the used temperature range.

 The successive coupling of the C-centres is predicted to be a kind of 
 hierarchical process, which leads to a non-Curie law in the magnetic susceptibility as, for example, 
 $\chi \propto T^{-\alpha}$. Such a process has been observed in P-doped Si with $\alpha \sim 0.6$ \cite{paa88,lak94}.
The FC susceptibility of our diamond samples in the as-received state follows a similar temperature dependence but
with $\alpha \simeq 0.35$,  indicating the superposition of a strong diamagnetic response, 
as expected if  a SC phase develops, see Fig.~\ref{t-dep}(a). 
The results in Fig.~\ref{t-dep}(a)  indicate also  that at high enough temperatures most of the 
C-centres 
depair and
contribute as  independent PM centres to the magnetic susceptibility.

 A  less conventional possibility follows from  arguments published in \cite{bas08} and \cite{mares08}: 
 spin-spin interaction is the most robust attractive interaction
 in this kind of doped systems, which may operate simultaneously to the electron-phonon interaction to create Cooper pairs. 
 It follows that, unlike  s-wave, p-wave  superconductors do not necessarily have their Cooper pairs mediated by phonons.
We may argue therefore that, instead of a s-wave pairing,  a p-wave SC state between the interacting donors could also be possible.
In this case the response to an applied magnetic field of a system of  p-wave symmetry Cooper pairs could produce
 a mix response of a SPM together to a  field hysteresis loop due to  the existence of vortices and/or
fluxons in the SC regions. 
As example, we refer to the theoretical study  of  the orbital magnetic dynamics in a p-wave superconductor 
with strong crystal-field anisotropy \cite{bra06}. In this case  the orbital moment of Cooper pairs (the directional order parameter) 
does not lead to a definite spontaneous magnetization, i.e. no hysteresis as in a SPM phase. 
Clearly, more experimental evidence tying the possible causes to a consistent, measurable effect is 
necessary. 

\subsection{What electron irradiation does and why the observed effect is of importance}

Let us now discuss the effect of the electron irradiation. One main result  is that  a nominal induced defect concentration $\sim 50$~ppm, 
of the order of the concentration of C-centres in the as-received state of the samples, 
 eliminates completely the anomalies in the magnetization. In would mean that 
the two phases, SC and the SPM phases, vanish simultaneously. 
Taking into account that most of those produced vacancies diffuse and/or give rise N-related 
defects, and the correlation found in \cite{bardia}, it is appropriate to correlate the vanishing of the anomalies after 
 irradiation with  the measured  decrease of $\sim 10\ldots 40$~ppm  of C-centres, see Table~I. 

Qualitatively, the irradiation effect we found is similar to the vanishing of SC by  He-ions irradiation
in B-doped diamond \cite{cre18}. This similarity 
is remarkable because  the studies in B-doped diamond  show  a suppression of SC after producing a vacancy concentration  
$\sim 2$\%, similar to the boron concentration.  This result also indicates that in the N-doped samples 
the SC/SPM regions should be
spread in the whole sample and not just at certain localized regions, each with orders of magnitude larger density of C-centres. 
If that were  the case, it would be
difficult to understand how a change  of $\sim 40$~ppm in the defect concentration (C-centres and/or other lattice defects) 
is enough to suppress the two phases. 
This experimental result supports the
notion that nitrogen  doping of diamond is extraordinary, especially because of the  low level of doping needed to
trigger the observed irreversibiities in the magnetization and at relatively high temperatures. 

The suppression of  SC 
in  B-doped diamond after He-irradiation has been explained assuming that the produced  
vacancies act as donors, which compensate the holes introduced
by the substitutional boron atom \cite{cre18}. Evidently this argument is not applicable 
in the case of the N-doped diamond, because nitrogen is already a donor in the diamond matrix.

From our results we may conclude that one  main reason for the vanishing of the responsible phases by 
electron irradiation is the  decrease by $\simeq 50\%$ of the average concentration of C-centres, see Table~I. 
However, would that  decrease in the C-centres concentration  be the only effect,  then we would expect to still 
observe the two phases, though with less
amplitude of the anomalies in the magnetization. This expectation is based on the rather weak dependence of
the $T_c$ on the average concentration of C-centres, obtained  from eight  diamond samples 
with  differences up to  a factor four \cite{bardia}.
It means that the irradiation-produced lattice defects  strongly affect the interaction 
between the remaining  C-centres.

The produced lattice defects through irradiation  (H1a-centres \cite{bab16}, remaining vacancies  \cite{cam00} and other
N-centres)  could affect not only the 
randomness of the lattice  strain, the distribution of the internal electric field and of the covalent bonds
that contribute to stabilize the SC phase \cite{liu20}, but also their magnetic moments could have
a  detrimental role in the coupling between the C-centres. If this would be the case, then a 
distribution of C-centres in a  diamond lattice with a similar amount of 
magnetic defects not necessarily would trigger SC but only PM. 
In fact, the susceptibility results {\em after irradiation}  indicate that most of the remaining C-centres contribute  independently 
to a PM state at all temperatures.

The studies in \cite{cre18} found that a partial recovery 
 of the SC occurs after annealing the  He-irradiated B-doped diamond samples.  A similar annealing
treatment done in  our samples also produces the  recovery of  the anomalies in the magnetization but their magnitude  
is smaller than in the as-received state  at similar temperatures,
see Fig.~\ref{fh}. Which is actually the effect of annealing?
 After electron irradiation, annealing at high temperatures and in vacuum reduces only slightly the  average concentration of 
C-centres ($\sim 6\%$, see Table~I) but   increases significantly the
homogeneity in their spatial distribution, see Fig.~\ref{CI-2}. Because the total amount
of C-centres did not change basically after annealing, see Table~I, the partial recovering of
the responsible phases after annealing, appears  to be related to the clear increase in  
homogeneity. Our IR results do not 
indicate that the concentration of H1a-centres decreases with  annealing.

\section{Conclusions}
Magnetization measurements of electron irradiated N-doped diamond crystals show the suppression of the anomalous  
irreversible behavior in applied magnetic field and temperature. The suppression occurs
after producing a decrease of a few tens of ppm in the  concentration  of   C-centres  measured  by IR absorption and EPR. 
This is remarkable because of the relatively low density of  C-centres, stressing the
extraordinary role of the C-centres in triggering those phenomena in diamond at relatively high temperatures.  Spectroscopy methods
like ARPES to get information on the changes in the band structure produced by nitrogen would be of high interest. However,
it is not clear whether such a technique would be successful using similar samples as we studied here, due to the small amount of mass
of the phases that originate the anomalies in the magnetization. 
We believe that future work should try to study the magnetic and electrical response locally in order to localize the regions of interest.
According to our results, the regions of interest should be all over distributed in the samples and therefore these local 
studies should have a reasonable chance of success. However, the main problem would be the large 
measuring scanning time (at a fixed temperature and magnetic field) for the typical areas of
the samples studied here.





\medskip
\textbf{Acknowledgements} \par 
The authors thank W. B\"ohlmann for the technical support. 
One of the authors (PDE) thanks G. Baskaran and G. Zhang for fruitful discussions on their work. 
The studies were supported by the DFG under the grant DFG-ES 86/29-1 and DFG-ME 1564/11-1. The work in Russia was partially funded by RFBR and NSFC research project 20-52-53051. 

\medskip

%

\end{document}